\RequirePackage{amsthm}

\documentclass[sn-mathphys,Numbered]{sn-jnl}

\usepackage{graphicx}%
\usepackage{multirow}%
\usepackage{amsmath,amssymb,amsfonts}%
\usepackage{amsthm}%
\usepackage{mathrsfs}%
\usepackage[title]{appendix}%
\usepackage{xcolor}%
\usepackage{textcomp}%
\usepackage{manyfoot}%
\usepackage{booktabs}%
\usepackage{algorithm}%
\usepackage{algorithmicx}%
\usepackage{algpseudocode}%
\usepackage{listings}%
\usepackage{siunitx}

\theoremstyle{thmstyleone}%
\theoremstyle{thmstyletwo}%

\theoremstyle{thmstylethree}%

\begin{document}

\title[A Hybrid Process for Integration of Organic Electrochemical Transistors for High Uniformity \& Reliability ]{A Hybrid Process for Integration of Organic Electrochemical Transistors for High Uniformity \& Reliability }


\author*[]{\fnm{Tommy} \sur{Meier*} \email{tommy.meier@tu-dresden.de}}
\author*[]{\fnm{Yeohoon} \sur{Yoon*} \email{yeohoon.yoon@tu-dresden.de}}
\author[]{\fnm{Laura} \sur{Teuerle}}
\author[]{\fnm{Ali} \sur{Solgi}}
\author[]{\fnm{Karl} \sur{Leo}}
\author[]{\fnm{Hans} \sur{Kleemann}}

\affil[]{\orgdiv{IAPP TU Dresden}, \orgname{Institute for Applied Physics, Technische Universit\"at Dresden}, \orgaddress{\street{N\"othnitzer Straße 61}, \postcode{01187}, \city{Dresden}, \state{Germany}}}


\abstract{Photolithography is believed to be a complementary technique to large-area printing, allowing for nanometer-scale integration and offering cost-efficiency. For organic electronics though, adapting photolithography is very challenging due to chemical incompatibilities. However, with the help of Alexander Zakhidov, orthogonal resins opened up the prospect of adapting the well-established process of photolithography for organic electronics. Here, we present a hybrid fabrication method for organic electrochemical transistors by combining orthogonal photolithography and inkjet printing, enabling high uniformity and reliability. We demonstrate how the resolution of each process affects the uniformity, and we explore the advantages of this process for device scaling and circuit integration.} 


%
\keywords{Hybrid, Organic, Microstructure, Electrical properties, Ink-jet printing}


\maketitle

\section*{Introduction}\label{sec1}

Organic electronics have experienced significant growth in recent years due to their advantages in low-cost, low-temperature processing, mechanical flexibility, and biocompatibility, allowing for a wide range of possible applications. In particular, organic semiconductors found their application in organic light-emitting diodes (OLEDs) for displays. In fact, OLEDs are nowadays the leading technology for mobile phone displays, and organic solar cells, transistors, sensors, and multifunctional devices are expected to enter the market in the near future. 

However, for a long time, this research field faced a significant challenge: the direct adaptation of photolithographic manufacturing techniques, which were key for the success of their inorganic counterparts. This limitation resulted from chemical incompatibilities between conventional photoresists and solvents, damaging the underlying organic layers \cite{lee2009orthogonal}. Although photolithography is considered to be a high-cost process, it is very established in semiconductor manufacturing, allowing for high yield and integration from the nanometer to micrometer range. In this regard, photolithography is considered to be an alternative patterning technique for organic electronics, complementing large-area and low-cost printing techniques for applications where high spatial resolution is needed.

A major step towards addressing this challenge of photolithography for organic electronics was presented by DeFranco \emph{et al.} in 2005 \cite{defranco2006photolithographic}. This method employed a protective parylene layer as an intermediary, which was subsequently removed through a lift-off process. Finally, an orthogonal patterning method based on hydrofluoroether solvents was pioneered in the group of Christopher Ober and with the help of Alexander Zakhidov, this technology has been applied to organic electronic devices in the following years \cite{zakhidov2008hydrofluoroethers, zakhidov2011orthogonal}. This breakthrough opened up the field of organic electronics to a well-known and established manufacturing process capable of achieving nanoscale resolutions. In fact, the patterning recipes developed by Alexander Zakhidov and the materials commercialized by Orthogonal Inc. have been adapted by many researchers for various types of organic devices, including transistors and OLEDs \cite{Nakayama2014, Kleemann2013, KLEEMANN2012, Sawatzki2018, Krotkus2014}.

Photolithography is the preferred patterning technique for thin-film devices in the inorganic semiconductor industry. However, in the field of organic electronics, cost-effective printing processes are typically preferred. In fact, organic semiconductors are ideal for such printing processes thanks to their low-temperature solution processability \cite{defranco2006photolithographic}. While screen printing may compromise resolution, it is ideal for rapid large-scale manufacturing. In contrast, inkjet printing offers precise digital control over print geometries by simultaneously minimizing material waste \cite{mattana2017inkjet}. 
However, with increasing spatial resolution or integration over multiple scales, the complexity and costs of these techniques rapidly increase, giving photolithography an edge over printing methods due to long-standing industrial process experiences and continuous optimizations.

\begin{figure}[h!]
\centering
\includegraphics[width=13.2cm]{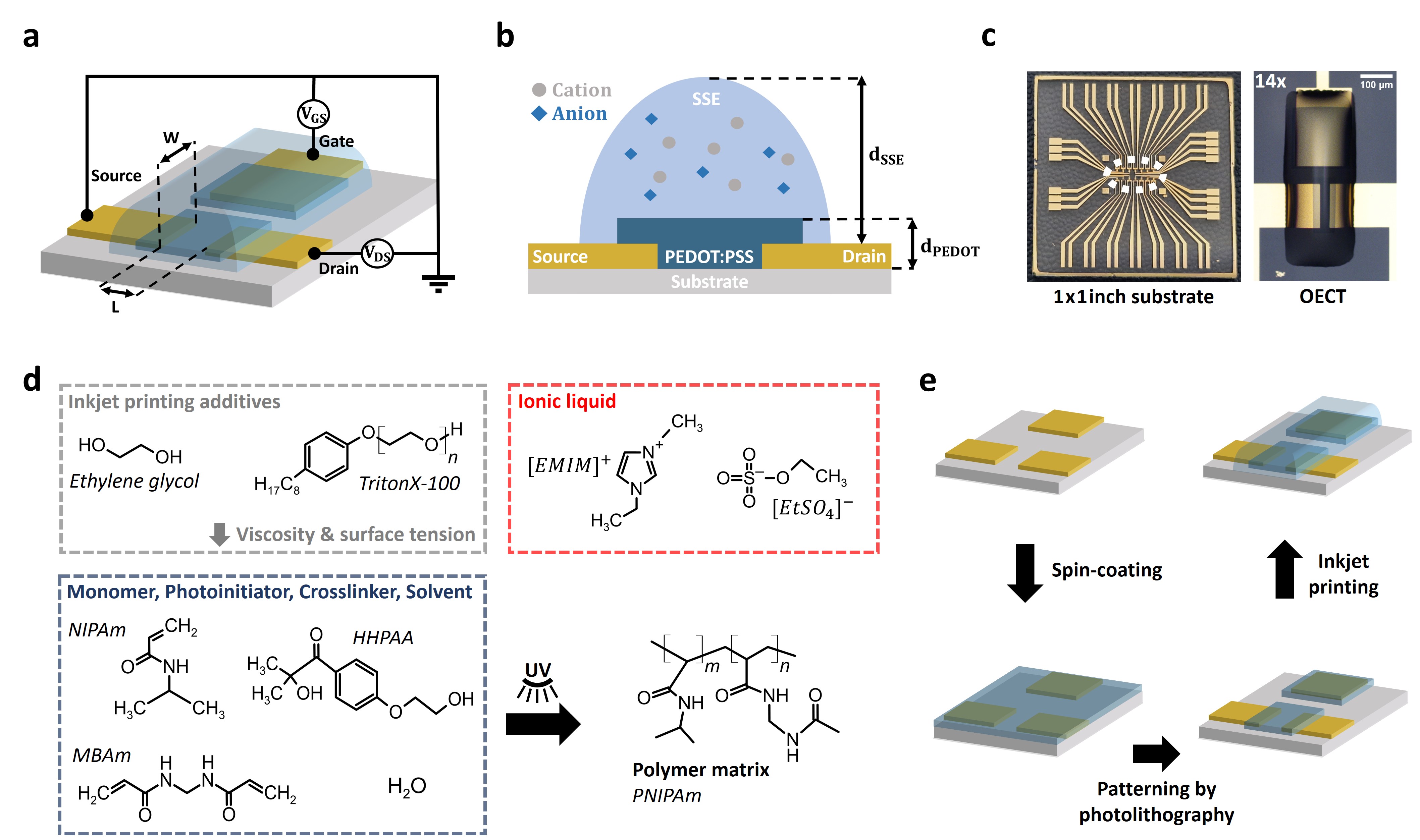}
\caption{\textbf{OECT structure and organic compounds} (\textbf{a}) 3D schematic and (\textbf{b}) cross-section of an OECT. (\textbf{c}) Image of a real layout where 14 OECTs are integrated onto a single substrate. (\textbf{d}) The SSE precursor solution is prepared by blending ionic liquid, monomer, photoinitiator, crosslinker, and solvent (cf. \cite{weissbach2022photopatternable}). To optimize it for the inkjet printing process, the viscosity and surface tension are decreased using ethylene glycol and TritonX-100, respectively. The crosslinked SSE structure is realized upon UV exposure. Details are given in the Methods section. (\textbf{e}) Schematic of the hybrid OECT fabrication process, which patterns the spin-coated PEDOT:PSS channel material via photolithography and applies the SSE through inkjet printing.}
\label{fig:fig1}
\end{figure}

A device where this effect of multi-scale integration becomes obvious is the organic electrochemical transistor (OECT). These 3-terminal devices are comprised of an organic mixed ionic-electronic conductor (OMIEC) that serves as the channel material and a gate electrode immersed in an electrolyte (see Figure \ref{fig:fig1}a, b). OECTs are experiencing increasing academic and industrial interest due to their ability to operate in liquid environments and their sensitivity to ions. Thanks to these unique properties, OECTs have been successfully implemented in bioelectronic applications. For instance, they have been used to record electrocardiogram (ECG) and electroencephalography (EEG) signals with a high signal-to-noise ratio \cite{leleux2015organic}, or as ion and enzymatic sensors \cite{strakosas2015organic, Tseng2021}. Furthermore, the capability of OECTs to exhibit synaptic short- or long-term plasticity positions them advantageously for neuromorphic applications \cite{Cucchi2021, van2017non}. Their seamless integration on large-area using screen or inkjet printing processes \cite{yang2022low} might enable the realization of intelligent circuits. However, as the transistor geometry $W$, $L$, $L_{ov}$, $d$ are the channel width, length, overlap and thickness, respectively (see Figure \ref{fig:fig1}a, b), determines the transconductance and saturation behaviour of OECTs \cite{bernards2007steady, Weissbach2022}, it is essential to control these geometry parameters with micrometer precision in order to ensure highest reliability and uniformity of transistor parameters, which is needed for circuit integration. In particular, we recently demonstrated that misalignment of the semiconductor layer and electrolyte on the length scale of \SI{5}{\micro\meter} can cause a loss of saturation of OECTs and threshold voltage shift with a channel length of more than \SI{100}{\micro\meter} \cite{Weissbach2022} - an effect that is detrimental for any kind of analog or digital circuit.

Here, we present a hybrid fabrication method for OECTs by combining orthogonal photolithography (as pioneered by Alexander Zakhidov) and inkjet printing, enabling high transistor uniformity alongside remarkable transfer reliability. In particular, we analyze how the resolution of each process influences the uniformity of performance, and we discuss the benefits of this hybrid process for device scaling and large-area integration of circuits. Finally, we showcase the application of the integrated OECTs using a photopatternable solid-state electrolyte (SSE) in a unipolar inverter setup. 




\section*{Results and Discussion}\label{sec2}

\subsection*{{\normalfont \textit{Fabrication Techniques}}}
We utilize the benchmark OMIEC poly(3,4-ethylenedioxythiophene) polystyrene sulfonate (PEDOT:PSS) \cite{rivnay2016structural} and an in-house developed solid-state electrolyte (cf. \cite{weissbach2022photopatternable}, Figure \ref{fig:fig1}d). However, the integration process can also be applied to more recently developed OMIECs, such as glycolated polythiophenes.

In the following, we investigate the impact of different fabrication methods on device performance. We fabricate OECTs using three distinct manufacturing techniques: a pure photolithography approach, a pure inkjet printing approach, and a hybrid method combining photolithographic patterning of the PEDOT:PSS channel and inkjet printing of the SSE (details of the processes are given in the Methods section). A schematic of the hybrid procedure is provided in Figure \ref{fig:fig1}e. For the sake of simplicity, the metal layer (Cr/Au for source, drain and gate in a side-gate configuration) is structured by photolithography and wet-etching for all devices. As we will show, the uniformity of the devices is mainly governed by the uniformity of the SSE and OMIEC. Consequently, the metal layer could also be done by printing without compromising the uniformity.

\begin{figure}[h!]
\centering
\includegraphics[width=13.2cm]{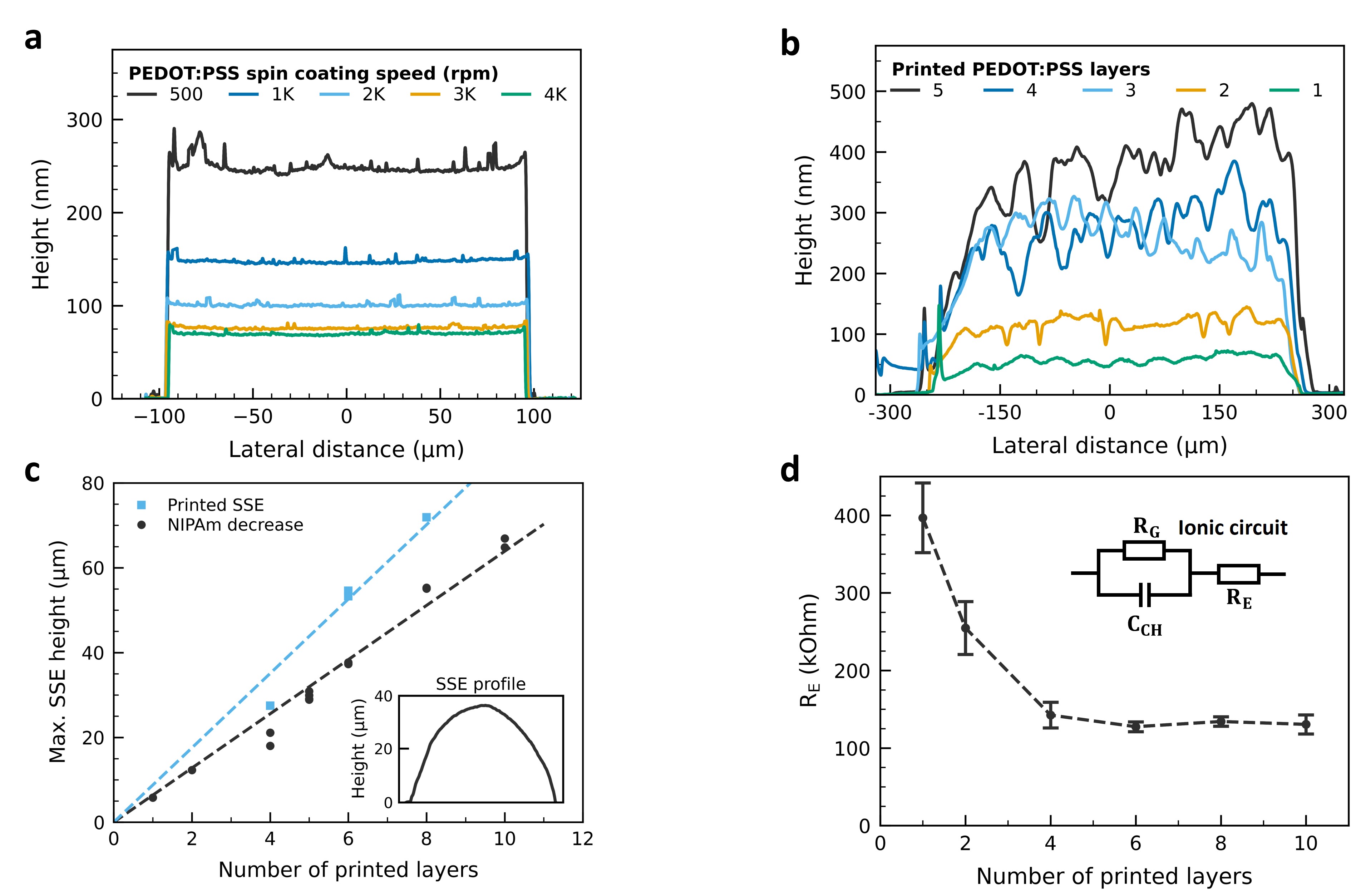}
\caption{\textbf{Influence of fabrication methods on PEDOT:PSS and SSE} Dependence of the PEDOT:PSS profile shape on (\textbf{a}) the lithographic fabrication technique with various spin-coating speeds, and (\textbf{b}) the inkjet printing method with 1\,-\,5 printed layers. (\textbf{c}) The peak height of droplet-shaped inkjet printed SSEs correlates linearly with the number of printed layers. However, as the concentration of NIPAm in the precursor solution is decreased, a corresponding decline in height is observed. (\textbf{d}) The electrolyte resistance is determined by fitting impedance spectroscopy data with the equivalent ionic R(CR)-circuit of the Bernards model \cite{bernards2007steady}. The dataset is associated with a hybrid OECTs with W\,/\,L ratio of 5.}
\label{fig:fig2}
\end{figure}

The structure of our devices is presented in Figure \ref{fig:fig1}c. Notably, a single substrate can accommodate up to 14 devices, which are all manufactured simultaneously. To assess the reproducibility of the transfer performance of these OECTs, we must examine the uniformity (thickness and dimensions) of the patterned functional layers first. Figure \ref{fig:fig2}a shows the height profiles of photolithographically patterned PEDOT:PSS. In this case, the channel thickness $d$ can be adjusted by altering the rotation rate during spin-coating. 
Using this technique, a consistently uniform and flat PEDOT:PSS layer can be achieved, and the thickness can be varied reliably from 70\,nm up to 250\,nm. Greater thicknesses are reached by consecutively repeating the coating process prior to the patterning step, which is based on orthogonal photoresist. The lateral dimensions of the PEDOT:PSS are defined by the lithography mask, development and etching process, yielding tolerances in the range of \SI{1}{\micro\meter}. In contrast, the maximum thickness of inkjet printed PEDOT:PSS is governed by the number of printed layers, as shown in Figure \ref{fig:fig2}b. While the printed layer profile closely aligns with the anticipated droplet shape, the non-uniformity and inconsistency across the examined profiles are clearly visible (although ink formulation and printing settings have been optimized). Several factors contribute to these shape inconsistencies. Firstly, the composition of the PEDOT:PSS solution differs based on the manufacturing process employed. A PEDOT:PSS dispersion with 5\,\% v/v ethylene glycol is utilized for the photolithography technique. This concentration is reported to provide optimal transistor performance while making a good compromise between ionic and electronic transport mechanisms \cite{rivnay2016structural}. In contrast, for inkjet printing a blend out of 1:1 PEDOT:PSS/H$_\text{2}$O+5\,\% DMSO + 1\,\% TritonX-100 is prepared. To meet specific ink requirements of the inkjet printer, the viscosity of pristine PEDOT:PSS is decreased with the addition of water and DMSO. Notably, DMSO, similar to ethylene glycol, enhances the conductivity of PEDOT:PSS via secondary doping \cite{ahmad2021mechanisms}. Furthermore, to optimize the surface tension, the surfactant TritonX-100 is introduced \cite{lopez2008characterization}. It is important to note that both DMSO and TritonX-100 can increase the jetting voltage, and previous research suggests that a higher voltage leads to faster droplet ejection, increased film roughness, and thickness \cite{garnett2005electrical}. Additionally, most solvents evaporate during spin-coating, resulting in even molecular distribution. In contrast, inkjet printing tends to deposit the solvent-rich PEDOT:PSS solution onto the substrate, which can lead to small clusters and uneven material distribution during the annealing step. Furthermore, due to the size of the droplets and the accuracy of the position of the print-head, the lateral dimension of the inkjet printed films can vary by up to $\sim$\,\SI{20}{\micro\meter}. Thus, the accuracy of height, length and width of the PEDOT:PSS layer is significantly better once orthogonal photolithography and dry-etching are used for the patterning rather than printing.
The second step of the integration process concerns the solid-state electrolyte. Again, we compare the profiles of the layers for the different integration processes. 
The profile of photolithographic patterned SSEs is similar to the shape of the PEDOT:PSS film due to the good solubility contrast of exposed vs unexposed areas. However, due to the high viscosity of the SSE, it is not suited for spin-coating. Instead, the SSE is drop-casted on the substrate, and a Teflon sheet covers the surface (see Methods section). However, this is a poorly controllable process, and the SSE height is limited to $\sim$\,\SI{17}{\micro\meter} by the viscosity. This lack of control leads to several micrometer-large variations in electrolyte height, even for devices produced simultaneously on the same substrate. In contrast, the number of printed SSE layers precisely determines the maximum SSE height, and a consistent direct proportionality is observed (see Figure \ref{fig:fig2}c). Reducing the concentration of the polymer monomers (NIPAm) in the precursor solution lowers the resulting SSE height and decreases the crosslinking efficiency. Furthermore, when comparing the shape of inkjet printed SSEs with printed PEDOT:PSS profiles, the former clearly offers a more uniform and smoother shape due to the high viscosity. 

To evaluate the electrolyte resistance, impedance electrochemical spectroscopy measurements are performed and fitted with the ionic R(CR)-circuit of the Bernards model \cite{bernards2007steady}. As presented in Figure \ref{fig:fig2}d, the resistance decreases rapidly with the SSE height. This decrease saturates after a minimum of 4 printed layers, corresponding to an SSE height exceeding \SI{20}{\micro\meter}. Consequently, SSE portions beyond these saturation heights no longer contribute to the electronic or ionic interactions within the OECT system. The reason for this saturation lies in the aspect ratio of the devices, where the height of the SSE becomes comparable to the lateral dimensions, e.g., the gap between the gate and source/drain. Furthermore, the deviations in electrolyte resistance are more pronounced for SSE heights below the saturation level. Throughout our experiments, we noted that printed SSEs require longer exposure times than their photolithography counterparts, and printing of just 1 or 2 layers results in insufficient crosslinking upon UV exposure. 

\subsection*{{\normalfont \textit{Influence of SSE height}}}
\begin{figure}[h!]
\centering
\includegraphics[width=13.2cm]{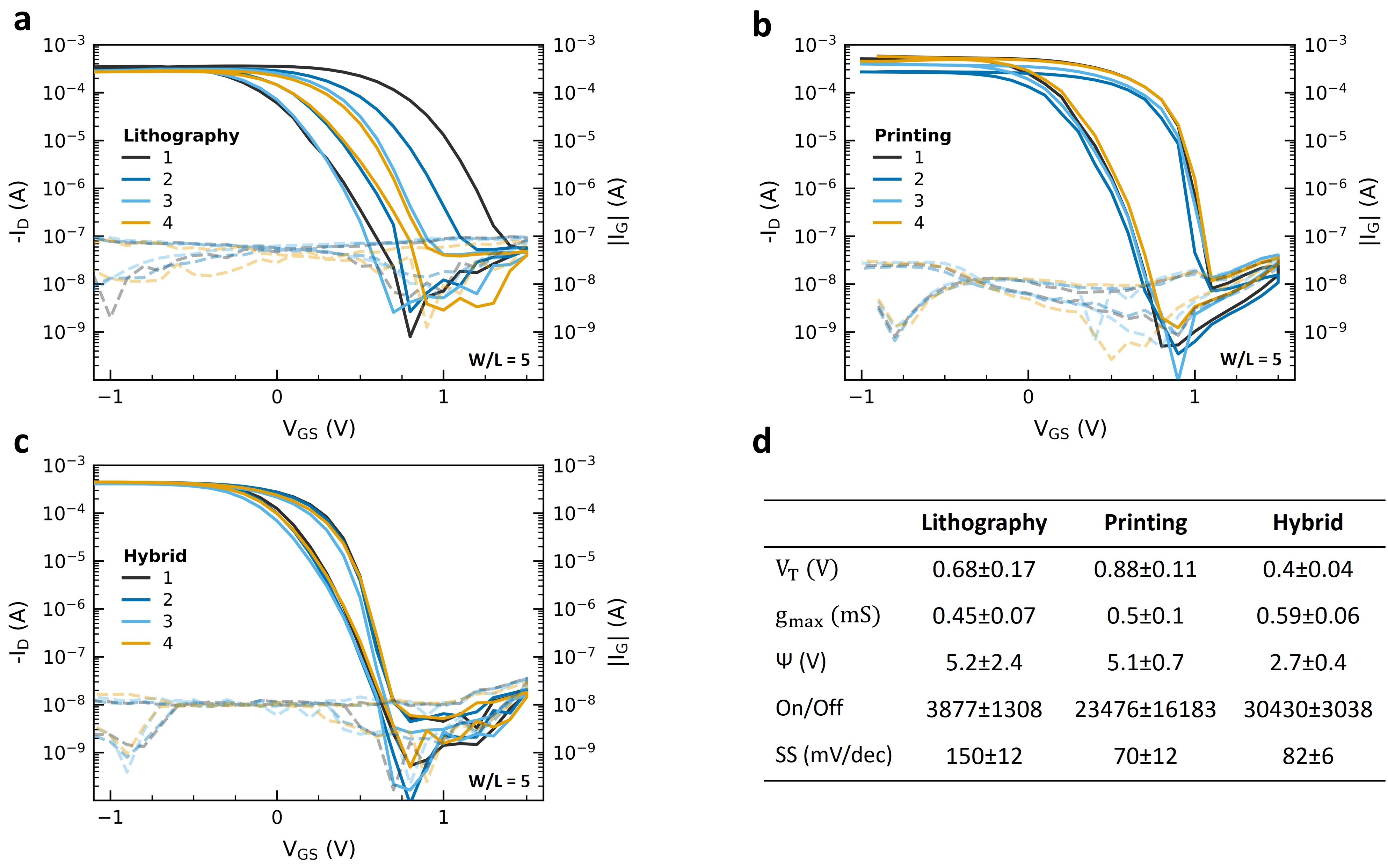}
\caption{\textbf{Transfer curve reliability} Comparison of transfer curves of four OECTs (numbered from 1 to 4) with a $W$\,/\,$L$ ratio of 5 across three different manufacturing processes: (\textbf{a}) lithography for both PEDOT:PSS (75\,nm) and SSE (inconsistent heights), (\textbf{b}) inkjet printing for both PEDOT:PSS (3 layers) and SSE (5 layers), and (\textbf{c}) the hybrid method combining lithography for PEDOT:PSS (75\,nm) and inkjet printing for SSE (5 layers). The source-drain voltage is -0.1\,V. (\textbf{d}) Corresponding comparison of characteristic transfer parameters (each dataset consists of 10 devices).}
\label{fig:fig3}
\end{figure}

Figure \ref{fig:fig3}a\,-\,c are comparing the transfer curves of photolithographically structured, fully printed, and hybrid OECTs, respectively. In order to determine the uniformity of performance more accurately, we extract parameters such as the threshold voltage, hysteresis strength ($\Psi$), maximum transconductance (g$_\text{max}$), subthreshold slope (SS) and on/off ratio from the transfer curves of 10 identical devices (see Figure \ref{fig:fig3}d). Photolithographically structured devices (Figure \ref{fig:fig3}a) show large variations in the threshold voltage (V$_{T}$) and hysteresis strength (determined by the parameter $\Psi$, cf. \cite{weissbach2022photopatternable}) among the 4 identical devices. Surprisingly, the transconductance is uniform despite the large differences in the threshold voltage.

The primary factor contributing to the non-uniformity of these devices is the variability in the SSE height from the lithography process. Figure \ref{fig:fig4}a and b clearly illustrate, that if the SSE height falls below \SI{20}{\micro\meter}, significant deviations in hysteresis strength and threshold voltage occur. Furthermore, it is worth noting the patterning of SSE using lithography tends to induce higher gate leakage currents to the SSE printing process (cf. Figures \ref{fig:fig3}a, b, c). This effect is due to residuals of the ionic liquid, which cannot be removed during the development of the SSE. 

In order to enhance reliability and uniformity, we adopt the inkjet printed SSE. As shown in Figures \ref{fig:fig3}b and c, significant improvements are seen in the uniformity using this process. In particular, the threshold voltage becomes almost independent of the SSE thickness if the thickness exceeds \SI{20}{\micro\meter}, which is very desirable as it lowers the requirements for process control. A detailed comparison between fully printed and hybrid devices shows that the hybrid OECTs give more uniform performance measures (see Figure \ref{fig:fig3}d), which is presumably due to the better uniformity of the PEDOT:PSS layers patterned by photolithography. Furthermore, hybrid OECTs reach significantly lower off-current values, leading to an improved transistor performance in terms of the on/off ratio. Interestingly, hybrid devices possess a much lower threshold voltage compared to lithography-made OECTs, which is beneficial for power saving in circuits.

In conclusion, when printing SSE, we achieve a very reliable threshold voltage, hysteresis strength, subthreshold swing, transconductance and on/off ratio. The main cause lies in the uniform printing of these thick layers and thickness-independent resistivity (see Figure \ref{fig:fig2}d). Therefore, the inkjet printing method for SSE fabrication proves to be a promising approach for enhancing device reliability as well as device performance. Furthermore, in comparison with the photolithographically patterned semiconductor, we reach the best uniformity due to a consistent PEDOT:PSS thickness compared to the printing process.


\subsection*{{\normalfont \textit{Impact of fabrication methods on PEDOT:PSS layer}}}

\begin{figure}[h!]
\centering
\includegraphics[width=13.2cm]{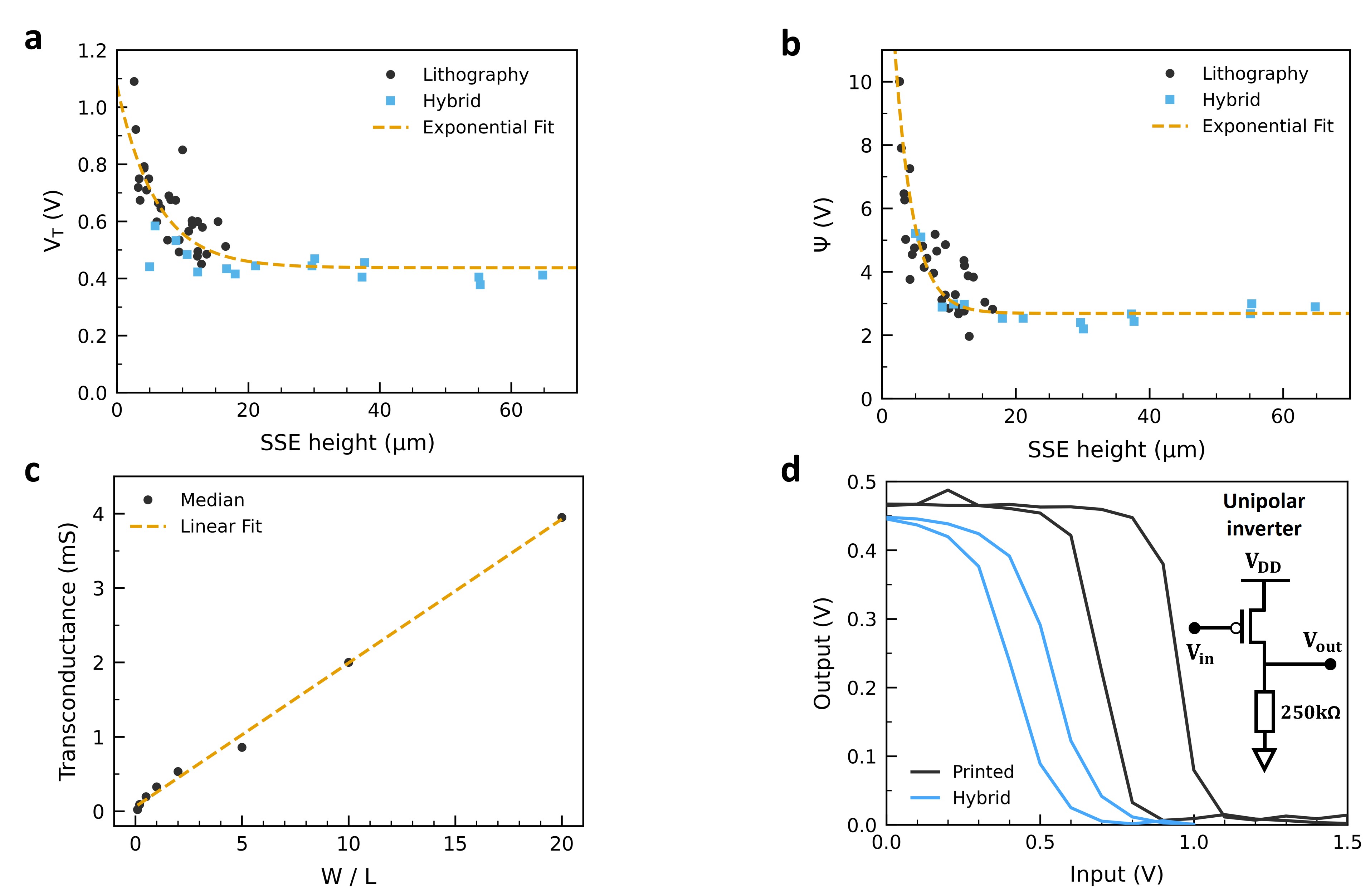}
\caption{\textbf{Transfer characteristics and unipolar inverter curve} Combined dataset for (\textbf{a}) the threshold voltage and (\textbf{b}) the hysteresis strength $\Psi$ (cf. \cite{weissbach2022photopatternable}) of OECTs fabricated with SSE using lithographic patterning and inkjet printing of 5 layers (both with a 75\,nm thick PEDOT:PSS channel patterned by lithography). There is an exponential-like dependence of these characteristic parameters on the SSE height. (\textbf{c}) Median transconductance values align well with the anticipated linear dependency on geometric scaling. (\textbf{d}) Comparison of unipolar inverters based on a fully printed and a hybrid OECT.}
\label{fig:fig4}
\end{figure}

By comparing the transfer curves in Figure \ref{fig:fig3}b and c and keeping in mind that the printed PEDOT:PSS is more than three times thicker than the spin-coated layer, we see that the transconductance of the OECTs is affected by the deposition method of the semiconductor. The Bernards model \cite{bernards2007steady} predicts a tight correlation between the channel current as well as the transconductance on the device geometry, which is an important factor for device scaling. Theoretically, fully printed OECTs should exhibit a higher transconductance compared to hybrid OECTs due to the channel thickness differences ($\sim$\,250\,nm for 3 printed layers of PEDOT:PSS and 75\,nm for 3000\,rpm spin-coated PEDOT:PSS). However, the obtained maximum transconductance of printed PEDOT:PSS OECTs is lower than that of hybrid OECTs (see Figure \ref{fig:fig3}d). This observation suggests that factors other than geometry may be contributing to these results. Therefore, it is essential to consider the hole mobility (\text{$\mu$}) and volumetric capacitance (C$^*$) of the channel material. Recent studies indicate that these two parameters are influenced by the microstructure of PEDOT:PSS, which will be greatly affected by the fabrication method and the two different PEDOT:PSS solution compositions used in this work \cite{kim2018influence}. The efficient movement of charge carriers depends on the degree of $\pi$-$\pi$ stacking in the semiconductor microstructure. Higher crystallinity results in a shorter d-spacing (010), indicating a more compact film structure with enhanced mobility \cite{sirringhaus1999two}. 
Additionally, it is important to achieve a uniform molecular-scale dispersion of ion-conductive polymers throughout the organic layer, as it allows for the efficient penetration of small ions throughout the channel \cite{rivnay2016structural}. While additional research is required to confirm whether the printed PEDOT:PSS exhibits less crystallinity, the factors mentioned above suggest that the polymer arrangement in the printed films may be less ordered in comparison to spin-coated films. In this regard, it has been reported that adding ethylene glycol to PEDOT:PSS results in a higher conductivity and \text{$\mu$}C$^*$ compared to formulations with added DMSO \cite{inal2017benchmarking}. Consequently, even though fully printed OECTs possess a thicker PEDOT:PSS film, they exhibit a relatively lower on-current and transconductance. Additionally, the non-uniform channel thickness of printed PEDOT:PSS films resulted in the largest observed transconductance deviation (see Figure \ref{fig:fig3}d).\\
Upon further comparison of the transfer curves among fully printed OECTs, significant differences in the on-current and, therefore, in the on/off ratio are observed. These inconsistencies arise from variations in film thickness, as shown in the height profiles in Figure \ref{fig:fig2}b). While fully printed OECTs exhibit a nearly consistent hysteresis, it is notably larger than that observed in hybrid OECTs. 
Thicker films may indeed have a higher likelihood of trapping ions compared to thin and crystalline films \cite{quill2021ion}. Therefore, ion trapping could result in huge hysteresis effects, impacting the transfer characteristics. The extent of trapping in thicker film can depend on various factors, including the specific material, the degree of disorder, and the film's morphology. However, the relationship between these factors and the observed hysteresis is complex, and further studies have to be performed to prove this relationship.\\
Despite these differences in the uniformity of device performance, we generally find that all OECTs fulfil the expected linear scaling \cite{rivnay2015high} of the transconductance with the geometry ratio $W$\,/\,$L$ (see Figure \ref{fig:fig4}c). While printed OECTs exhibit better device-to-device reliability than OECTs solely relying on the photolithography process, further optimization of the PEDOT:PSS ink recipe and printing parameters might even reduce the variability further.

While having the same amount of printed SSE layers, fully printed OECTs exhibit a larger threshold voltage than their hybrid counterparts. This phenomenon can potentially be explained by an ionic circuit model \cite{tseng2023threshold, Weissbach2022}. As we have shown previously, the high impedance of the electrochemical double-layer capacitance formed at the semiconductor-electrolyte interface makes the electrolyte resistance of a 5-layer SSE ($\sim$\,110k$\Omega$) negligible compared to the imaginary part of the impedance \cite{Weissbach2022}.
Consequently, we can obtain an equivalent circuit from two capacitors connected in series, describing the double-layer capacitance at the gate and OMIEC interface. The ionic charge accumulated at the gate capacitor is equal to the charge accumulated in the channel capacitor (but with the opposite sign), and we can formulate the following equation \cite{Weissbach2022}.
\begin{equation}
    C_{Gate}(V_{GS} - V_{eff})  = C_{Ch}(V_{el} - V_{Ch})
\label{cap1}
\end{equation}

Here, $C_{Gate}$ and $C_{Ch}$ represent the capacitance of the gate and channel, $V_{GS}$ is the applied gates-source voltage, $V_{el}$ represents the constant voltage within the electrolyte, and $V_{Ch}$ denotes the channel potential. Note that in this work, we are utilizing a capacitive PEDOT:PSS gate electrode, which is patterned during the same step as the channel material. Although generally not allowed, we can set the channel potential to zero as the drain-source voltage in our study is much smaller than the gate-source and threshold voltage. Consequently, we are assuming a constant potential along the direction of the transistor channel. Using this simplification, we express Equation \ref{cap1} in terms of V$_{el}$ and it becomes:
\begin{equation}
   V_{el}  = \frac{C_{Gate} V_{GS}}{C_{Gate}+C_{Ch}} 
\label{cap2}
\end{equation}

If the channel capacitance is one order of magnitude smaller than the gate capacitance, as seen in Equation \ref{cap2}, the effective voltage V$_{el}$ applied to the channel-electrolyte interface becomes equal to the gate-source voltage (V$_{GS}$). However, if the channel thickness increases, V$_{el}$ decreases. The equation suggests a reduced voltage drop across the channel, leading to an increase in threshold voltage. Consequently, it is assumed that thicker printed PEDOT:PSS films are leading to higher threshold voltages compared to hybrid OECTs. Additionally, variations in film thickness are considered to contribute to the threshold voltage variability among printed devices. 

\subsection*{{\normalfont \textit{Unipolar Inverter}}}
To showcase the application of our OECTs, we integrated a fully printed OECT and a hybrid OECT in a unipolar inverter setup (see Figure \ref{fig:fig4}d). Thereby, the OECTs are connected to a passive resistor of 250\,k$\Omega$, which serves as a pull-down device. The specific selection of the resistor was based on the channel resistance at an input voltage of 0.25\,V. If the p-type OECT is in the on-state, the voltage drop will occur across the passive resistor, resulting in an output voltage equal to the supply voltage V$_{DD}$, which was set to 0.5\,V. Conversely, if the device is in the off-state by applying a high input voltage V$_{in}$, the voltage drop will occur across the device, and the output voltage is zero.\\
Since the transconductance difference between the two OECT types is not significant, the voltage gains were nearly identical, with values of 1.8 (V/V) for an inverter based on a hybrid OECT and 1.9 (V/V) for the fully printed OECT inverter. However, depletion-mode OECTs are generally unfavourable for circuitry because of high power dissipation. Therefore, research efforts are focused on lowering the threshold voltage through various methods, including chemical and structural modifications \cite{tseng2023threshold,keene2020enhancement}. However, our results show that the right choice of the fabrication technique is crucial for achieving this objective. Consequently, due to their reduced threshold voltage, hybrid OECTs consume less power than inverters based on fully printed OECTs, and their lower variability might be essential for circuit integration.

\section*{Conclusion}\label{sec3}
In this study, we utilize printing and lithography techniques to fabricate OECTs with excellent uniformity and reliability in their electrical properties. We analyzed the uniformity of the PEDOT:PSS and SSE layers and unveiled that PEDOT:PSS patterned via lithography exhibits a consistently uniform surface and thickness, unlike their printed counterparts. Furthermore, the microstructure of the printed semiconductor layer affects the on-current level and transconductance of OECTs. However, adjusting our printing setup by optimizing the ink formulation and print parameters might help to enhance the reproducibility of fully printed devices in the future. In contrast to the PEDOT:PSS layer, only the inkjet printed SSE consistently achieved reproducible heights with a smooth droplet-shaped profile.\\
Our hybrid OECT successfully combines the two most consistent fabrication techniques for PEDOT:PSS and SSE. This process yields OECTs with a remarkable degree of uniformity in the transfer curve, high on/off ratios ($>$\,10$^{4}$), and low threshold voltages of $\sim$ 0.4\,V. This fabrication procedure provides a good foundation for scalable, reliable, and consistent devices suitable for integrated circuits.\\
Moreover, incorporating an SSE in our hybrid OECT system revealed a rapidly decreasing dependence of the threshold voltage and hysteresis on the SSE height. Optimal device performance is observed in the parameter saturation regime for SSE heights surpassing \SI{20}{\micro\meter}. Last but not least, we successfully implemented our hybrid OECT in an unipolar inverter setup with a gain of 1.8 (V/V).

These observations pave the way for the fabrication of OECTs with reproducible and specific transfer characteristics and have potential applications in analog/digital circuits or in neuromorphic circuits, where hysteresis tuning can achieve different short- or long-term memory effects. 

\section*{Methods}\label{sec4}

\bmhead{Substrate preparation} A 1\,inch x 1\,inch glass substrate with a 3\,nm Cr and a 50\,nm Au layer was cleaned in acetone by ultrasonication for 15\,min, followed by spin-coating (3000\,rpm for 60 s) of the positive photoresist AZ\,1518 and an annealing step for 60\,s at 110\,°C. UV exposure was done using a photomask with the SÜSS Microtec MJB4 mask aligner system (SÜSS Microtec AG, Germany, I-line 365\,nm, lamp power 167\,W) for 12\,s. The exposed photoresist areas were removed with the AZ\,726 MIF developer, followed by a gold and chromium wet-etching process. Subsequently, the substrate was cleaned in acetone by ultrasonication for 15\,min to remove the unexposed photoresist and was treated with O$_\text{2}$-plasma for 5\,min.
It is worth mentioning that fully printed OECTs used a different channel geometry (\SI{250}{\micro\meter}\,x\,\SI{50}{\micro\meter}) in contrast to the geometry used in lithography and hybrid devices (\SI{150}{\micro\meter}\,x\,\SI{30}{\micro\meter}). However, the characteristic $W$\,/\,$L$ ratio was maintained at 5 to ensure comparability with the other manufacturing processes.

\bmhead{Photolithographic patterning of PEDOT:PSS} A 1\,\% aqueous PEDOT:PSS dispersion (Clevios PH1000) with 5\,\%v/v of ethylene glycol was spin-coated (3000\,rpm for 60\,s) and annealed for 20\,min at 120\,°C. The substrate was then spin-coated (3000\,rpm for 60\,s) with the negative orthogonal photoresist OSCoR\,4020, followed by pre-baking for 1\,min at 103\,°C. UV exposure took place for 14\,s using the SÜSS mask aligner. After post-baking for 1 \,min at 103\,°C, the unexposed photoresist was removed with orthogonal developer 103a, and the PEDOT:PSS was etched away in a Diener electronic ATTO plasma-cleaner (Diener electronic GmbH \& Co. KG, Germany) using O$_\text{2}$- and Ar-plasma. The remaining photoresist was removed with orthogonal stripper 900, followed by ultrasonic cleaning in ethanol for 15\,min.

\bmhead{Photolithographic patterning of SSE} The SSE precursor solution consists out of 1\,ml of water, 1.5\,ml of ionic liquid (EMIM-EtSO$_\text{4}$), 750\,mg of the monomer N-isopropylacrylamide (NIPAm), 200\,mg of the crosslinker N,N’-Methylenebis(acrylamide) (MBAm), and 20\,mg of the photoinitiator 2-hydroxy-4’-(2-hydroxyethoxy)-2-methylpropiophenone (HHPAA). Additionally, an adhesion promoter was prepared by mixing 9\,ml ethanol, \SI{90}{\micro\liter} Silane A174, and \SI{180}{\micro\liter} concentrated acetic acid. The PEDOT:PSS patterned substrate was immersed in the promoter solution for 10\,min at 50\,°C, followed by cleaning with ethanol and annealing for 10\,min at 100\,°C. The SSE precursor solution was drop casted on the substrate and covered with a Teflon foil to spread the solution. UV exposure was done for 20\,s using the SÜSS mask aligner, and the excess unexposed solution was removed afterwards by blowing with N$_\text{2}$.

\bmhead{Inkjet printing of PEDOT:PSS} The ink consists out of a 1:1 volume ratio of PEDOT:PSS (Clevios PH1000) and water, together with 5\,\%(v/v) DMSO and 1\,\%(v/v) TritonX-100. For filtration of the ink a PVDF \SI{0.45}{\micro\meter} filter is used. A Dimatix Materials DMP-2800 inkjet printer
(Fujifilm Dimatix Inc., USA) with a piezoelectric DoD 12-nozzle ‘Samba’ cartridge was utilized. The average drop volume of this cartridge type is 2.4\, pL, resulting in a drop radius of $\sim$\,\SI{8}{\micro\meter}. The optimum drop performance is achieved if the ink has a viscosity in the range of 4\,-\,8\,mPa·s, a surface tension of 28\,-\,32\,mN/m and a higher boiling point than 100\,°C. For printing a drop spacing of \SI{15}{\micro\meter} is used. The PEDOT:PSS ink was printed on the substrate in the desired geometry, followed by annealing at 120\,$^\circ$C for 10\,min.

\bmhead{Inkjet printing of SSE} The SSE precursor solution of the photolithographic patterning process was modified by adding 1.5\,ml of ethylene glycol and 1\,\%v/v of TritonX-100 (see Figure \ref{fig:fig1}d) and filtered with a PVDF \SI{0.45}{\micro\meter} filter. The adhesion promoter step from the photolithographic patterning method was performed, and the modified SSE solution was printed with the Dimatix inkjet printer on the transistor area with multiple layers, followed by UV exposure for 120\,s.

\bmhead{Electrical characterization} Transfer characteristics were evaluated within a nitrogen-filled glovebox using two Keithley 236 SMUs controlled by the SweepMe! software (sweep-me.net). For all measurements, the source-drain voltage was maintained at -0.1$\,$V and each gate voltage step was set to 0.1$\,$V with a hold time of 0.2$\,$s in between.

\bmhead{Impedance spectroscopy} Impedance measurements were performed within a nitrogen-filled glovebox using a Metrohm Autolab PGSTAT302N potentiostat/galvanostat (Metrohm AG). The impedance spectroscopy was realized by measuring between the shorted source-drain electrodes and the gate electrode with a root mean square amplitude of 10\,mV in the potentiostat mode. Fitting of this data was done with the software ZView$^{\copyright}$ by applying an equivalent R(CR)-circuit model.

\bmhead{Surface investigation} To obtain the height profile of PEDOT:PSS and the SSE, a Veeco Dektak 150 surface profiler (Veeco Instruments Inc., USA) with a \SI{5}{\micro\meter} radius stylus was
utilized. The measurement was conducted in the standard scan mode with a measurement range of \SI{524}{\micro\meter} and a stylus force of 3\,mg.


\vspace*{10mm}
\bmhead{Author Contribution}
T.M. and Y.Y. carried out the experimental work, including device fabrication and measurements. L.T. and A.S. co-developed the inks and optimized the process for uniformity. K.L. and H.K. supervised the work and designed the experiments. T.M., Y.Y. and H.K. wrote the manuscript. 
\bmhead{Funding}
    T.M., Y.Y., L.T. and H.K. are grateful for funding provided by the European Commission through the project BAYFLEX (101099555). A.S. and H.K. thank the Bundesministerium für Bildung und Forschung (BMBF) for funding from the project BAYOEN (01IS21089). Furthermore, H.K. is grateful for funding from the German Research Foundation (DFG) under the E-Mask project (KL 2961/5-1).
\bmhead{Conflict of Interest}
The authors declare no conflict of interest.
\bmhead{Data Availability}
Further data is available from the authors upon reasonable request.

\end{document}